\documentclass[prl,aps,twocolumn]{revtex4}

\usepackage{amssymb}
\usepackage{amsmath}
\usepackage{epsfig}

\begin{document}
\title{Mixed Qubit Cannot Be Universally Broadcast}
\author{Lin Chen}
\author{Yi-Xin Chen}
\affiliation{Zhejiang Insitute of Modern Physics, Zhejiang
University, Hangzhou 310027, China}

\begin{abstract}
We show that there does not exist any universal quantum cloning
machine that can broadcast an arbitrary mixed qubit with a
constant fidelity. Based on this result, we investigate the
dependent quantum cloner in the sense that some parameter of the
input qubit $\rho_s(\theta,\omega,\lambda)$ is regarded as
constant in the fidelity. For the case of constant $\omega$, we
establish the $1\rightarrow2$ optimal symmetric dependent cloner
with a fidelity 1/2. It is also shown that the $1\rightarrow M$
optimal quantum cloning machine for pure qubits is also optimal
for mixed qubits, when $\lambda$ is the unique parameter in the
fidelity. For general $N\rightarrow M$ broadcasting of mixed
qubits, the situation is very different.
\end{abstract}
\maketitle

Quantum mechanics poses many restrictions differing with the rule
of classical world. For example, the superposition principle
prohibits the perfect cloning of arbitrary unknown quantum state
(no-cloning theorem) \cite{Zurek}. However, it has been shown that
the optimal cloning, namely the construction of a quantum cloning
machine (QCM) with the best fidelity \cite{Jozsa} of clone at the
output is achievable by many authors
\cite{Buzek,Bruss,Gisin,Werner,Niu}. In existing literatures, they
are called universal QCM (UQCM) in the sense that the fidelity is
independent of the input state \cite{Scarani}, and the UQCMs are
further divided into the symmetric and asymmetric kinds according
to that whether the fidelity of clones are identical or not.
Remarkably, all these results are based on the pure inputs, yet
little is known about the quantum cloner whose input state is
mixed. Here we raise a fundamental problem as follows: given one
arbitrary mixed qubit state and certain ancilla state as inputs,
is there a (symmetric or asymmetric) UQCM that can transform them
into $M$ copies with (identical or nonidentical) constant
fidelities? This question should be the first to be considered for
the mixed states cloning, just like the case of treating the pure
inputs. To our knowledge, no systematic result on this topic has
been obtained in the existing literature. On the other hand, mixed
states cloning is more useful than pure states cloning in
experiments, due to the decoherence and noise of the environment.
The rapid progress of quantum cryptography \cite{Curty} also
requires a deeper understanding of the mixed states cloning.

Generally, a mixed qubit state can be written as
\begin{equation}
\rho_s\equiv\lambda|\psi\rangle\langle\psi|+(1-\lambda)|\psi^{\bot}\rangle\langle\psi^{\bot}|,
\end{equation}
where a pure qubit state has the form $
\left|\psi\right\rangle=\cos\theta\left|0\right\rangle+e^{i\omega}\sin\theta
\left|1\right\rangle$ and its orthogonal state is $
|\psi^{\bot}\rangle=-\sin\theta\left|0\right\rangle+e^{i\omega}\cos\theta
\left|1\right\rangle$, defined in the computational basis
$\{\left|0\right\rangle,\left|1\right\rangle\}$. Every qubit has a
corresponding point in the Bloch sphere \cite{Chuang}, and the
three parameters $\theta,\omega$, and $\lambda$ in $\rho_s$
explicitly determine the position of this point. Specially,
$\rho_s$ becomes a pure state when $\lambda=0$ or 1 (a pure qubit
is always on the sphere and vice versa). So the set of pure qubits
actually belongs to that of mixed qubits, and the latter has a
more abundant content in the aspect of cloning. It was pointed out
firstly by \cite{Barnum} that the mixed qubits can be $broadcast$,
since every observer can obtain the identical local state by
tracing over other systems in the joint state at the output of
QCM. So the broadcasting is a generalization of cloning, and we
also refer to the tensor product of identical states as the mixed
states cloning \cite{Barnum,Ariano}.

The introduction of a new parameter $\lambda$ makes it difficult
to find out a UQCM for mixed qubits. We recall the celebrated
conclusion \cite{Barnum} which asserts two mixed states can be
broadcast if and only if (iff) they are commuting. Such a
commutativity makes it possible to provide the $a$ $priori$
information (the simultaneous eigenstates) of initial state. In
fact, one can broadcast any mixed state when the a priori
information is enough, even if they are in a noncommuting set. For
example, given the content of $\theta$ and $\omega$ in $\rho_s$ as
input, one can perform a unitary operator $U$ on a copy of
$\rho_s$ and the ancilla $|0\rangle^{\otimes M-1}$ such that
$U|\psi\rangle|0\rangle^{\otimes{M-1}}=|\psi\rangle^{\otimes M},
U|\psi^{\bot}\rangle|0\rangle^{\otimes{M-1}}=|\psi^{\bot}\rangle^{\otimes
M}$. Then by tracing over $M-1$ systems in the output state, we
readily realize the broadcasting of the mixed qubits. As it can be
immediately generalized to a state of arbitrary dimension, we say
that the a priori information plays an important role in
broadcasting mixed states.

In this paper, we shall show that there does not exist any UQCM
for mixed qubits, if there is no a priori information of initial
state provided for the QCM. This explains why the shrinking factor
\cite{Bruss} cannot generally represent the merit of QCM for mixed
qubits. Then we catalog the $dependent$ QCMs, whose fidelity
contains some parameter of the initial state $\rho_s$, by whether
this parameter is known by the cloner. We first obtain the
$1\rightarrow2$ optimal symmetric dependent QCM with a fidelity
1/2, when the content of $\omega$ is provided. An important case
arises when only the parameter $\lambda$ exist in the fidelity but
we know nothing about $\rho_s$, since this QCM relates to the
existing results of pure states cloner. In this case, we will show
that the $1\rightarrow M$ optimal UQCM for pure qubits
\cite{Gisin} is also optimal for mixed qubits. We thus have given
a general result of dependent QCM on broadcasting mixed states.

Now, we start the derivation of our results. We evaluate the merit
of broadcasting by the fidelity
$F(\rho_1,\rho_2)=\Big[\mbox{tr}\sqrt{\sqrt{\rho_1}\rho_2\sqrt{\rho_1}}\Big]^2
$, between two mixed qubits $\rho_1$ and $\rho_2$. Suppose one of
the output state is $\rho_A\equiv\left(\begin{array}{cc}
x & y \\
y^{*} & 1-x
\end{array}\right),
$ then the fidelity is
\begin{eqnarray}
F(\rho_A,\rho_s)&=&\frac12+\frac12(2\lambda-1)(2x-1)\cos2\theta\nonumber\\
&+&\frac12(2\lambda-1)(e^{i\omega}y+e^{-i\omega}y^*)\sin2\theta\nonumber\\
&+&2\sqrt{(\lambda-\lambda^2)(x-x^2-|y|^2)}.
\end{eqnarray}
The definition of UQCM requires the fidelity is independent of all
parameters of the initial state, and under this constraint we
investigate the universality of QCM for mixed qubits. The most
general method of broadcasting quantum states can be described by
a unitary operator (QCM) acting on the system of initial copies
and ancilla, and then tracing out all but one subsystems of the
output state \cite{Barnum}. Here we are supplied with one copy of
$\rho_s$ as input, and the ancilla system is in a $d-$dimensional
state $\rho_{anc}$. So the unitary operator $U$ is $2d\times2d$,
and one output copy is in the state
\begin{equation}
\rho_A=\mbox{tr}_B[U(\rho_s\otimes\rho_{anc})U^\dag],
\end{equation}
where we regard $B$ as the residual system. Since no information
of the initial state is obtainable, both $U$ and $\rho_{anc}$ are
constant. We can thus write the output state as a linear
combination of two density operators containing only two
parameters $\theta$ and $\omega$, i.e.,
$\rho_A=\lambda\rho_1(\theta,\omega)+(1-\lambda)\rho_2(\theta,\omega)$.
Let the entries of $\rho_A$ be $x\equiv A\lambda+B,y\equiv
C\lambda+D$, then the coefficients $A,B,C,D$ are only the
functions of $\theta$ and $\omega$. Without loss of generality, we
set the auxiliary state in a diagonal form by absorbing its
diagonalizing similarity matrix into the QCM, that is,
$\rho_{anc}=\mbox{diag}(c_0,...,c_{d-1})$.

To obtain the accurate expression of the output state for the
calculation of the fidelity, we partition the cloner as
$U=(U^{2d\times d}_0,U^{2d\times d}_1)$, where a further partition
is
\begin{equation}
U^{2d\times d}_i\equiv\left(\begin{array}{ccc}
|u^i_{0,0}\rangle & \cdots & |u^i_{0,d-1}\rangle \\
\vdots& \ddots & \vdots\\
|u^i_{2^M-1,0}\rangle & \cdots & |u^i_{2^M-1,d-1}\rangle
\end{array}\right),i=0,1.
\end{equation}
Here, $|u^i_{j,k}\rangle$ are constant vectors with the dimension
$2^{1-M}d,j=0,...,2^M-1,k=0,...,d-1$. This partition always holds
true since the expectant QCM broadcasts the initial qubit to M
copies. By tracing over all but one systems, we obtain the output
copy $\rho_A$ in terms of the entries $x$ and $y$, where the
coefficients satisfy
\begin{eqnarray}
A+2B=\sum^{d-1}_{i=0}\sum^{2^M-1}_{k=0,even}c_i
(\langle u^0_{k,i}|u^0_{k,i}\rangle+\langle u^1_{k,i}|u^1_{k,i}\rangle),\nonumber\\
C+2D=\sum^{d-1}_{i=0}\sum^{2^M-1}_{k=0,even}c_i (\langle
u^0_{k+1,i}|u^0_{k,i}\rangle+\langle
u^1_{k+1,i}|u^1_{k,i}\rangle).
\end{eqnarray}
Here ``even" means the sum happens only when $k$ is even.
Evidently, both the sums are constant, and we rename them
$A+2B=E_x,C+2D=E_y$. After replacing $x$ and $y$ by these
parameters in the fidelity, the UQCM requires
$\left.F(\rho_A,\rho_s)\right|_{\lambda=0}=\left.F(\rho_A,\rho_s)\right|_{\lambda=1}$.
It yields
\begin{equation}
(2E_x-2)\cos2\theta+e^{i\omega}E_y\sin2\theta+e^{-i\omega}E^*_y\sin2\theta=0.
\end{equation}
To keep this equation as an identity, it holds that $E_x=1$,
$E_y=0$, and it is thus easy to verify
$F(\rho_A,\rho_s)|_{\lambda=1/2}=1.$ This evidently contradicts
with the no-cloning theorem. The conclusion can be depicted as
follows.

\textit{No-Universality Theorem} (NUT). Given an arbitrary mixed
qubit and any auxiliary state as input, it is impossible to create
a (symmetric or asymmetric) QCM copying this qubit universally, if
no a priori information of this qubit is
provided.\hspace*{\fill}$\blacksquare$

The theorem asserts that we cannot find out a UQCM when the set of
input states extends to mixed qubits. This is the essential
difference between the QCMs for only pure qubits and for all
qubits. We notice that the shrinking factor $f$ \cite{Bruss} for
the input $\rho_s$ and the output
$\rho_A=\frac{1-f}{2}I^{2\times2}+f\rho_s$, can uniquely evaluate
the quality of QCM when the input state is pure, for there is a
linear relation between the fidelity and $f$. However it does not
hold for the scenario of mixed states. According to the NUT, the
fidelity must be dependent on some parameter of the initial state.
Because the parameter affects the merit of QCM, the shrinking
factor is no more the unique variant in the fidelity. In addition,
one can easily verify that the fidelity $F(\rho_A,\rho_s)$ is no
more a linear function of $f$. In this case, the shrinking factor
will not generally be the figure of merit for QCM, when we study
the problem of broadcasting mixed states.

The systematic method for broadcasting mixed states should be to
find out series of dependent QCMs (DQCM); that is, one can in
advance consider some parameter of the input state as a
``constant", such that the fidelity is the function of only this
parameter. We should distinguish two kinds of dependent QCMs: the
one with the a priori information of the parameter, e.g.,
$U(\theta,\omega,\lambda)$, $\rho_{anc}(\theta,\omega,\lambda)$
and the other without it. For simplicity we call them DQCM-I and
DQCM-II respectively. Evidently, the quality of DQCM-I is not less
than that of DQCM-II no matter which parameter is regarded as
constant. In the existing literature such as the phase covariant
QCM \cite{Cinchetti} and the real cloner \cite{Navez} choosing the
pure states $|\psi\rangle$ as inputs, the authors indeed evaluated
the parameters with values such that $\theta=\pi/4$ and
$\omega=0$, respectively, so they all belong to DQCM-I. As far as
we know, the contributions to quantum cloning so far pay little
attention to the difference of these two kinds of DQCMs.
Specially, the existing cloner for mixed states did not clarify it
at all \cite{Ariano,Fan2}. We stress that, the a priori
information indeed plays an essential role in the dependent
quantum cloning, which can be seen in the following text. So one
has to consider both the cases of DQCM-I and II when constructing
the dependent cloner.

Due to the complicated mathematics, there is so far little result
on the DQCMs even for pure states. However sequentially, we will
show some basic results of DQCMs broadcasting mixed states, by in
turn choosing the parameter $\omega$ and $\lambda$ as the unique
constant in the fidelity. We suppose the input state is arbitrary
mixed qubit $\rho_s$, and the constant parameter is obtainable in
the case of DQCM-I. In addition, such a cloner cannot be perfect,
since it is known that when $\omega=0$, the $1\rightarrow2$
optimal fidelity is $\frac12+\frac{1}{\sqrt2}$ by the real cloner
\cite{Navez}, and the optimal QCM \cite{Bruss} gives out 5/6 when
$\lambda=0$. Both of them are special cases in our cloner.

We first consider the parameter $\omega$ constant in the
$1\rightarrow2$ cloner. Based on the proof of NUT, we are going to
find out all possible values of constant fidelity. Let
$E_y=|E_y|e^{i\delta}$. Recall equation (6) is necessary in the
present case, we have $E_x=1, |E_y|\cos(\delta+\omega)=0$.
Following the proof of NUT, one can easily find out the choice of
$E_y=0$ leads to a perfect cloning of $\rho_s$. For the other case
of $\cos(\delta+\omega)=0$, we need find out the expressions of
$B$ and $D$ in the clone $\rho_A$, by tracing over the residual
systems of the output state. One can check that $B=\langle
L_0|L_0\rangle$ and $D=\langle L_1|L_0\rangle$, where the two
$d^2\times1$ vectors read
\begin{eqnarray}
|L_0\rangle=\left(\begin{array}{c}
\sqrt{c_0}|w_{0,0}\rangle \\
\sqrt{c_0}|w_{2,0}\rangle \\
\vdots \\
\sqrt{c_{d-1}}|w_{0,d-1}\rangle \\
\sqrt{c_{d-1}}|w_{2,d-1}\rangle \\
\end{array}\right),
|L_1\rangle=\left(\begin{array}{c}
\sqrt{c_0}|w_{1,0}\rangle \\
\sqrt{c_0}|w_{3,0}\rangle \\
\vdots \\
\sqrt{c_{d-1}}|w_{1,d-1}\rangle \\
\sqrt{c_{d-1}}|w_{3,d-1}\rangle \\
\end{array}\right),\nonumber\\
|w_{i,j}\rangle=-\sin\theta|u^0_{i,j}\rangle+e^{i\omega}\cos\theta
|u^1_{i,j}\rangle,\hspace{85pt}\nonumber\\
i=0,1,2,3,j=0,...,d-1.\hspace{125pt}
\end{eqnarray}
As $\lambda$ should not be a variant in the fidelity, we deduce
that $x-x^2-|y|^2$ can be exactly divided by $\lambda-\lambda^2$;
that is, $\langle L_0|L_0\rangle\langle L_1|L_1\rangle=|\langle
L_1|L_0\rangle|^2$.

However due to the Schwarz inequality, this equation holds iff
$|L_0\rangle=g(\theta,\omega)|L_1\rangle$. The orthogonality of
columns of the unitary operator $U$ then yields
\begin{equation}
(g(\pi/2,\omega)^*g(0,\omega)+1)(\langle
u^0_{1,i}|u^1_{1,i}\rangle+\langle u^0_{3,i}|u^1_{3,i}\rangle)=0,
\end{equation}
for the case of $c_i>0$. It is not hard to derive $E_y=0$ if the
equation $g(\pi/2,\omega)^*g(0,\omega)+1=0$ holds true, which
implies the fidelity is 1 again. For the case of $\langle
u^0_{1,i}|u^1_{1,i}\rangle+\langle u^0_{3,i}|u^1_{3,i}\rangle=0,$
we employ the trick that the fidelity is invariant under the
average of $F(\rho_A(\theta),\rho_s(\theta))$ and
$F(\rho_A(-\theta),\rho_s(-\theta))$, and it leads to
\begin{equation}
F(\rho_A,\rho_s)|_{\lambda=0}=\frac12+\frac a2\cos^22\theta,
\end{equation}
where $a$ is constant. As $\theta$ is a variant, it is precluded
from the fidelity iff $a=0$. Thus the DQCM can be universal with a
single-copy fidelity equaling $1/2$ or $1$, if the a priori
information of $\omega$ is obtainable. In addition, such a quantum
cloner must be DQCM-I, since $\cos(\delta+\omega)=0$ always holds
no matter the quantum cloner is symmetric (with one fidelity
equaling 1/2 or 1 with the change of $\omega$) or asymmetric (with
two different fidelities equaling 1/2 or 1 for the two copies
respectively). This means that the other kind of dependent cloner
does not exist. So the a priori information is essential in this
case.

We have found that the symmetric cloner actually exist, e.g., when
$d=4$ the DQCM can be $U^{8\times8}(\omega)$, whose first and
fifth columns are
$(e^{i(-2\omega+\pi)},0,e^{i(-\omega+\pi/2)},0,e^{i(-\omega+\pi/2)},0,1,0)^T$
and
$(0,e^{i(-2\omega+\pi)},0,e^{i(-\omega+\pi/2)},0,e^{i(-\omega+\pi/2)},0,1)^T$,
and the rest of columns can be obtained by the unitarity of DQCM.
Let the ancilla be in the state $|0\rangle$, then the output copy
is in a pure state
$(e^{i(-\omega+\pi/2)}|0\rangle+|1\rangle)/\sqrt2$. The fidelity
of the clone equals 1/2, since it permits arbitrary mixed qubit to
be the input. It is lower than that of the real cloner
\cite{Navez}, which takes the state
$\cos\theta|0\rangle+\sin\theta|1\rangle$ as input. In addition,
there is no other symmetric DQCM whose fidelity of clones can be
either $1/2$ or $1$ with the change of $\omega$. This can be
derived from the fact that no $1\rightarrow2$ cloner can
explicitly duplicate the set of qubit $\rho_s$ with a definite
$\omega$. To account for this fact, one simply deduce that
$B(\theta=0)=0,B(\theta=\pi/2)=1$, which is required by the
condition of explicit cloning $E_x=1,E_y=0$. On the other hand, as
the QCM is symmetric, all calculations hold true in the case of
$|u^0_{1,i}\rangle\leftrightarrow|u^0_{2,i}\rangle,
|u^1_{1,i}\rangle\leftrightarrow|u^1_{2,i}\rangle,i=0,...,d-1.$ It
is easy to show $B=\sin^2\theta,D=0$, which leads to a
contradiction of
$F(\rho_A,\rho_s)|_{\lambda=0}=\frac{1+\cos^22\theta}{2}=1$, while
$\theta$ is a variant. So our cloner is the optimal symmetric
cloner with a priori information of $\omega$. One may apply the
present technique to the case of constant $\theta$, but there
indeed exists the perfect QCM duplicating the set of $\rho_s$ with
$\theta=0,\pi/2$ since it is a commuting set \cite{Barnum}. This
greatly enlarges the region of constant fidelity, so it becomes
difficult to determine the form of symmetric QCM.

Let us move to analyze the case of regarding $\lambda$ as
constant, and we will work with the $1\rightarrow M$ DQCM-II. That
is, the unitary operator $U$ and the auxiliary state contain no
$\lambda$. We notice that when $\lambda=0$ or $1$, the mixed qubit
$\rho_s$ becomes a pure state, whose optimal cloning has been
broad investigated \cite{Buzek,Bruss,Gisin,Werner}. In particular,
it is pointed out by \cite{Bruss} that if the cloner treats all
input pure qubit in the same way, then the Bloch vector of the
initial state can only be rescaled due to the hairy theorem. This
yields
\begin{equation}
|\psi\rangle\langle\psi|\rightarrow
z|\psi\rangle\langle\psi|+(1-z)|\psi^\perp\rangle\langle\psi^\perp|,
\end{equation}
where $z\in[0,1]$ is a constant. Since the pure qubits are a part
of the mixed qubits, this result also applies to the latter. The
process of the transformation itself reads
\begin{eqnarray}
\rho_s&=&\lambda|\psi\rangle\langle\psi|+(1-\lambda)|\psi^\perp\rangle\langle\psi^\perp|\rightarrow\nonumber\\
\rho_A&=&\lambda(\ z|\psi\rangle\langle\psi|+(1-z)|\psi^\perp\rangle\langle\psi^\perp|\ )\nonumber\\
&+&(1-\lambda)(\
z|\psi^\perp\rangle\langle\psi^\perp|+(1-z)|\psi\rangle\langle\psi|\
).
\end{eqnarray}
So we obtain the form of the output copy that keeps the fidelity
relates to $\lambda$ only, namely
\begin{eqnarray}
F(\rho_A,\rho_s)&=&[\sqrt{(\lambda(1-z)+(1-\lambda) z)(1-\lambda)}\nonumber\\
&+&\sqrt{(\lambda z+(1-\lambda)(1-z))\lambda}\ ]^2.
\end{eqnarray}
The constant $z$ is bounded by using the $1\rightarrow M$ optimal
cloner for pure qubits \cite{Gisin}, i.e.,
$F(\rho_A,\rho_s)|_{\lambda=0}=z\leq\frac{2M+1}{3M}$. Notice
$F(\rho_A,\rho_s)$ is monotonically increasing with $z$, thus we
reach the optimal fidelity when $z=\frac{2M+1}{3M}$; that is,
\begin{eqnarray}
F(\rho_A,\rho_s)&=&\bigg[\sqrt{\Big(\lambda\frac{M-1}{3M}+(1-\lambda)\frac{2M+1}{3M}\Big)(1-\lambda)}\nonumber\\
&+&\sqrt{\Big(\lambda\frac{2M+1}{3M}+(1-\lambda)\frac{M-1}{3M}\Big)\lambda}
\ \bigg]^2.
\end{eqnarray}
It is the upper bound of the fidelity under the constraint of
DQCM-II with constant $\lambda$. Fortunately, we have found that
the $1\rightarrow M$ optimal QCM \cite{Gisin} reaches this bound,
yet we take the mixed state $\rho_s$ as input here. So we indeed
propose a new kind of optimal dependent quantum cloner for mixed
states.

Let us make some observations about the optimal fidelity. First,
it saturates the bound of fidelity for optimal pure cloner, and it
also implies an explicit broadcasting for the case of
$\lambda=1/2$. Second, the fidelity is symmetric around the middle
point $\lambda=1/2$, i.e.,
$F(\rho_A,\rho_s)|_{\lambda}=F(\rho_A,\rho_s)|_{1-\lambda}$, which
coincides with the fact that $\rho_s(\lambda)$ and
$\rho_s(1-\lambda)$ represent the same sphere surface.
Furthermore, $F(\rho_A,\rho_s)$ is monotonically increasing with
$\lambda\in[0,1/2]$. This means that the optimal QCM can duplicate
the mixed qubit $\rho_s$ with a better effect as its Bloch vector
shrinks, and the fidelity becomes 1 corresponding to the maximally
mixed state. One can understand this fact as follows. When the end
of the Bloch vector of one qubit approaches the center of the
Bloch sphere, we are increasing the "mixedness" of the qubit. In
other words, the input state $\rho_s$ becomes less distinguishable
as we shorten the length of its Bloch vector $|2\lambda-1|$, and
thus the qubits with this length become close to each other.
Evidently, a concrete cloner can always improve its efficiency
when the input states have less difference. Finally, we observe
the length of $\rho_A$'s Bloch vector is
$|2\lambda-1|\cdot|2z-1|\leq|2\lambda-1|$, which implies the
optimal QCM is not only shrinking the output qubit for the pure
inputs \cite{Gisin}, but also doing that for the mixed inputs. We
doubt that whether it is possible to ``lengthen" the output qubit
by other QCMs, e.g., in the case of DQCM-I with a constant
$\lambda$.

We briefly investigate the $N\rightarrow M$ quantum cloner for
mixed qubits. The first question to be solved is whether there
exist a UQCM copying $\rho^{\otimes N}_s$ with a constant
fidelity. This is a task with more initial resource, and we cannot
immediately extend the NUT of $1\rightarrow M$ broadcasting mixed
qubits to this case. Nevertheless, there is the proof implying
that such a cloner is also dependent \cite{Ariano}. Second as
shown in the text, the shrinking factor $f$ does not generally
reflect the merit of the cloner for mixed states, so one cannot
directly establish the optimal $N\rightarrow M$ cloner by
maximizing $f$ \cite{Fan2}. For example, if one follows the
procedure for $1\rightarrow M$ cloner for mixed qubits and obtain
the optimal fidelity
$F(\rho_A,\rho_s)|_{\lambda=0}=z=\frac{NM+M+N}{M(N+2)}$ by
choosing $f=\frac{N}{M}\frac{M+2}{N+2}$, which is the upper bound
of shrinking factor by Gisin's optimal QCM \cite{Gisin}, the
expression (12) can reach 1 iff when $\lambda=1/2$. However, the
latest progress in \cite{Ariano} has shown that there exist
perfect dependent cloner with the initial states $\rho^{\otimes
N}_s$, even when $\rho_s$ is not maximally mixed. This implies
that the $N\rightarrow M$ cloner has different characters compared
to the case in the present paper.

In conclusion, we have proven the no-universality theorem for the
case of $1\rightarrow M$ broadcasting mixed qubits. Then we
investigated two special dependent QCMs such that one parameter
$\omega$ and $\lambda$ is regarded as constant in the fidelity,
respectively. For the former case, we have provided the optimal
symmetric cloner with the fidelity 1/2, and it is shown that the
$1\rightarrow M$ optimal QCM for pure qubits is also optimal in
the latter case. Our results are helpful to explore the dependent
quantum cloner for mixed qubits.

The work was partly supported by the NNSF of China Grant
No.90503009 and 973 Program Grant No.2005CB724508.

\end{document}